\documentclass[%
 reprint,
 superscriptaddress,
 showpacs,preprintnumbers,
 amsmath,amssymb,
 aps,
 prb,
 longbibliography,
]{revtex4-1}

\usepackage{graphicx}
\usepackage{dcolumn}
\usepackage{bm}
\usepackage{color}
\usepackage{ulem}

\begin{document}

\preprint{APS/123-QED}

\title{
Loop Liquid in an Ising-Spin Kondo Lattice Model on a Kagome Lattice
}

\author{Hiroaki Ishizuka}
\affiliation{
Department of Applied Physics, University of Tokyo, Hongo, 7-3-1, Bunkyo, Tokyo 113-8656, Japan
}

\author{Yukitoshi Motome}%
\affiliation{
Department of Applied Physics, University of Tokyo, Hongo, 7-3-1, Bunkyo, Tokyo 113-8656, Japan
}

\date{\today}

\begin{abstract}
Phase diagram of an Ising-spin Kondo lattice model on a kagome lattice is investigated by a Monte Carlo simulation.
We find that the system exhibits a peculiar ferrimagnetic state at a finite temperature, in which each triangle is in a two-up one-down spin configuration but the spin correlation does not develop any superstructure. 
We call this state the loop liquid, as it is characterized by the emergent degree of freedom, self-avoiding up-spin loops.
We elucidate that the system shows phase transitions from the loop liquid to ferrimagnetically ordered states and a crossover to a partially ferromagnetic state by changing the electron density and temperature. 
These can be viewed as crystallization and cohesion of the loops, respectively.
We demonstrate that the loop formation is observed in the optical conductivity as a characteristic resonant peak.
\end{abstract}

\pacs{
75.10.Kt, 
71.10.Fd, 
05.10.Ln  
}
\maketitle

The interplay between charge and spin degrees of freedom in electrons has long been studied as one of the central problems in condensed matter physics. 
In particular, itinerant electron systems coupled to localized spins are fundamental models in the study of such interplay in correlated electrons.
In these systems, effective spin-spin interactions mediated by itinerant electrons play a crucial role in the magnetism.
For instance, spin-glass behavior in metallic alloys with doped magnetic ions is driven by the so-called Ruderman-Kittel-Kasuya-Yosida interactions~\cite{Ruderman1954,Kasuya1956,Yosida1957}.
Another example is metallic ferromagnetic (FM) behavior in perovskite manganese oxides, which is understood by the double-exchange (DE) interaction~\cite{Zener1951,Anderson1955}.
On the other hand, the interplay between charge and spin also triggers significant changes of the electronic structure and transport properties.
In the perovskite magnanese oxides, the competition between the FM DE and antiferromagnetic (AFM) super-exchange interactions plays an important role in the colossal magneto-resistance~\cite{Tokura2000}. 

The interest in the spin-charge coupling has recently been extended to frustrated systems, in which geometrical frustration provides an additional degree of freedom for controlling the system.
In these systems, a simple AFM ordering of localized spins is suppressed by geometrical frustration, and instead, a disordered spin state with strong local correlations is expected to be realized at a sufficiently low temperature ($T$).
Such a ``liquid-like" spin state is anticipated to have characteristic effects on the coupled itinerant electrons.
Indeed, the importance of characteristic noncoplanar spin textures was discussed for an unconventional anomalous Hall effect and peculiar metallic behavior observed in Nd$_2$Mo$_2$O$_7$~\cite{Taguchi2001} and Pr$_2$Ir$_2$O$_7$~\cite{Nakatsuji2006,Machida2007}.

Stimulated by such interesting experiments, many theoretical studies have been done recently. 
For instance, characteristic metal-insulator transitions were found in extended Falicov-Kimball models with local constraint on the spatial configuration of localized particles~\cite{Ishizuka2011}. 
A similar problem was also studied for the model with localized spins~\cite{Jaubert2012}.
Meanwhile, a noncoplanar spin correlation in an Ising-spin Kondo lattice model on a pyrochlore lattice was studied for explaining the resistivity minimum in Pr$_2$Ir$_2$O$_7$~\cite{Udagawa2012,Chern2012-1}. 
A similar noncoplanar correlation on a kagome lattice was shown to induce a charge gap and quantum anomalous Hall effect~\cite{Ishizuka2013,Chern2012-2}.
These results indicate that liquid-like states can be realized in frustrated spin-charge coupled systems and their peculiar spin correlations significantly affect the electronic and transport properties.

\begin{figure}
   \includegraphics[width=\linewidth]{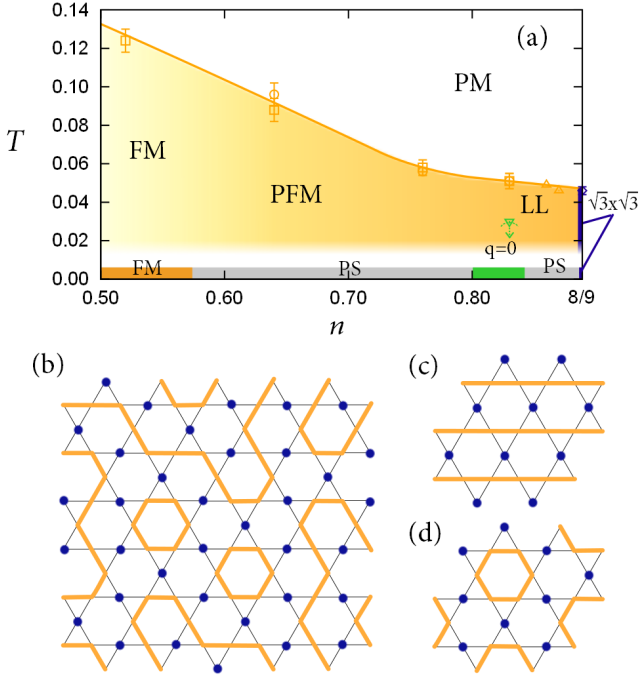}
   \caption{(color online).
   (a) Phase diagram of the model in Eq.~(\ref{eq:H}) at $J=6$ obtained by the Monte Carlo simulation.
   The symbols shows the critical temperatures $T_c$ for magnetic states: ferromagnetic (FM), partially ferromagnetic (PFM), loop liquid (LL), $q=0$ ferrimagnetic ($q=0$), and $\sqrt3\times \sqrt3$ ferrimagnetic ($\sqrt3\times \sqrt3$) states.
   $T_c$ for the $\sqrt3\times \sqrt3$ state at $n=8/9$ is shown by the diamond, while the upper limit for $T_c$ for $q=0$ state $n=0.83$ is shown by the downward triangle, which is given by the temperature we reached with $N_{\rm s}=8^2$ calculations.
   The squares (circles) show $T_c$ determined from the Binder analysis of $m$ ($P$), and the upward triangles show $T_c$ determined by the system-size extrapolation of the peak of $\chi_{m}$.
   The curve connecting the symbols is a guide for the eyes.
   The strip at the bottom is the ground state phase diagram obtained by the variational calculation for three magnetic orders, FM, $q=0$, and $\sqrt3\times \sqrt3$. 
   PS is the phase separation between the neighboring two phases.
   The schematic pictures of the magnetic states are given for (b) LL, (c) $q=0$, and (d) $\sqrt3\times \sqrt3$.
   The bold lines denote the loops connecting up-spin sites and the dots show down-spin sites.
   }
   \label{fig:diagram}
\end{figure}

In this study, to explore such liquid states and exotic electronic properties, we investigate a Kondo lattice model with Ising localized moments on a two-dimensional kagome lattice.
By using a Monte Carlo (MC) method, we show that the model exhibits a locally-correlated spin state with a fractional magnetic moment.
We call this ferrimagnetic (FR) liquid-like state the loop liquid, as it can be viewed as a soup of self-avoiding up-spin loops mixed with isolated down spins [see Fig.~\ref{fig:diagram}(b)]. 
Although the electronic structure was previously studied by assuming similar loop liquids~\cite{Jaubert2012}, our results provide a convincing evidence of the thermodynamic stability of the loop liquid for the first time to our knowledge.
The obtained phase diagram includes FM, partially FM, $q=0$ and $\sqrt{3}\times\sqrt{3}$ FR states in addition to the loop liquid; the phase transitions and crossover are interpreted as crystallization and cohesion in terms of the loops, respectively.
We also demonstrate that the optical conductivity develops characteristic peaks corresponding to the formation of the self-avoiding loops.

We consider a single-band Kondo lattice model on a kagome lattice with localized Ising spin moments.
The Hamiltonian is given by
\begin{eqnarray}
H = -t \! \sum_{\langle i,j \rangle, \sigma} \! ( c^\dagger_{i\sigma} c_{j\sigma} + \text{H.c.} ) + J \sum_{i}\sigma_i^z S_i.
\label{eq:H}
\end{eqnarray}
The first term represents hopping of itinerant electrons, where $c_{i\sigma}$ ($c^\dagger_{i\sigma}$) is the annihilation (creation) operator of an itinerant electron with spin $\sigma= \uparrow, \downarrow$ at the $i$th site, and $t$ is the transfer integral.
The sum $\langle i,j \rangle$ is taken over nearest-neighbor (NN) sites on the kagome lattice.
The second term is the on-site interaction between localized spins and itinerant electrons, where $\sigma_i^z = c_{i\uparrow}^\dagger c_{i\uparrow}-c_{i\downarrow}^\dagger c_{i\downarrow}$ represents the $z$ component of the itinerant electron spin and $S_i = \pm 1$ denotes the localized Ising spin at the $i$th site; $J$ is the coupling constant (the sign of $J$ does not matter in the present model). 
Hereafter, we take $t=1$ as the unit of energy, the lattice constant $a = 1$, the Boltzmann constant $k_{\rm B} = 1$, and $e^2/h$ as the unit of conductance.

Thermodynamic properties of the model in Eq.~(\ref{eq:H}) are studied by a MC simulation which is widely used for similar models~\cite{Yunoki1998}.
The calculations were conducted up to the system size $N=3\times N_{\rm s}$ with $N_{\rm s}=9^2$ under the periodic boundary conditions ($N_{\rm s}$ is the number of three-site unit cells).
To deal with the freezing of MC sampling, some of the low-$T$ data were calculated starting from a mixed initial spin configuration of low-$T$ ordered and high-$T$ disordered states~\cite{Ozeki2003}.
The thermal averages were calculated for typically 15000-80000 MC steps after 5000-18000 MC steps for thermalization.
In addition, the ground state phase diagram is also obtained by comparing the energy of the dominant phases found in the MC simulation~\cite{Ishizuka2013b}.

\begin{figure}
   \includegraphics[width=\linewidth]{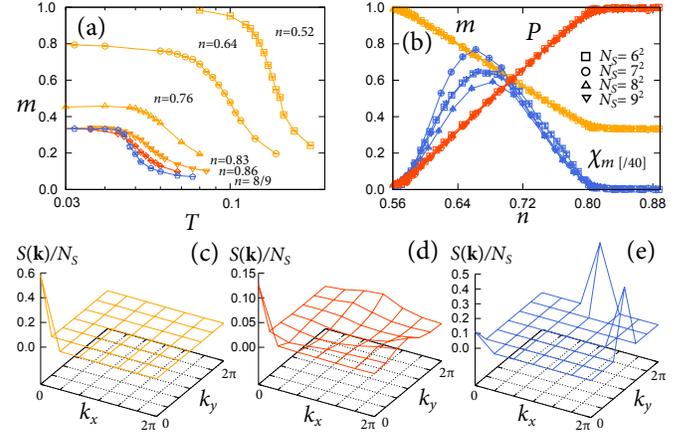}
   \caption{(color online).
   (a) MC results for $T$ dependences of $m$ at different $n$. 
   The data at $n=8/9$ are calculated for $N_{\rm s}=9^2$, while the others for $N_{\rm s}=8^2$.
   (b) $n$ dependences of $m$, $\chi_m$, and $P$ at $T=0.03$ for $N_{\rm s}=6^2$, $7^2$, $8^2$, and $9^2$.
   The MC results of $S({\bf k})/N_{\rm s}$ are shown for (c) $n=0.65$, (d) $n=0.84$, and (e) $n=8/9$ at $T=0.03$ and $N_{\rm s}=9^2$.
   }
   \label{fig:ndata}
\end{figure}

Figure~\ref{fig:diagram}(a) shows the phase diagram obtained by the MC simulation at $J=6$ while varying electron density $n=\sum_{i\sigma} \langle c_{i\sigma}^\dagger c_{i\sigma} \rangle/N$.
As lowering $T$, the system exhibits a phase transition with developing a net magnetization $m = \sqrt{\langle(\sum_i S_i / N)^2\rangle}$. 
$T$ dependences of $m$ are shown in Fig.~\ref{fig:ndata}(a). 
In the low density region for $n \lesssim 0.56$, $m$ approaches its saturated value $1$ in the low-$T$ limit, namely, the system exhibits a fully-polarized FM order. 
This phase is connected to the FM phase in the the large $J$ region, which is induced by the DE mechanism~\cite{Zener1951,Anderson1955}. 
While increasing $n$, the low-$T$ value of $m$ decreases from $1$ and continuously becomes smaller as $n$ becomes larger, as shown in Figs.~\ref{fig:ndata}(a) and \ref{fig:ndata}(b). 
At the same time, the probability to find a two-up one-down spin configuration in each triangle, $P=\sqrt{\langle(\sum_{\nu} 3p_{\nu}/2N)^2\rangle}$, increases continuously from zero [Fig.~\ref{fig:ndata}(b)]; here, $p_{\nu}=1 (-1)$ for two-up one-down (one-up two-down) and otherwise $p_{\nu}=0$, and the sum is over all triangles.
The spin structure factor $S({\bf k})$ for the same sublattice is featureless except for the peak at ${\bf k}={\bf 0}$, as shown in Fig.~\ref{fig:ndata}(c); here, $S({\bf k}) =  \frac1{N_{\rm s}} \sum_{i,j \in \alpha} \langle S_i S_j \rangle \exp\left({\rm i}{\bf k}\cdot{\bf r}_{ij}\right)$, where ${\bf r}_{ij}$ is the vector from $i$th to $j$th site, and the sum is taken for the sites $i,j$ in the same sublattice $\alpha$.
We call this region with the reduced $m$ the partially ferromagnetic (PFM) phase~\cite{note_pfm}. 

In the region of $0.8 \lesssim n < 8/9$, however, the low-$T$ value of $m$ becomes almost independent of $n$, and saturates to a fractional value $m=1/3$, as shown in Fig.~\ref{fig:ndata}(a). 
In this region, most of the triangles on the kagome lattice are in two-up one-down spin configurations, namely, $P \simeq 1$ [Fig.~\ref{fig:ndata}(b)]. 
As shown in Fig.~\ref{fig:ndata}(d), $S({\bf k})$ does not show any sharp peak except for the one at ${\bf k}={\bf 0}$, indicating that this state has no superstructure.
Hence, this FR state is a peculiar Coulombic state subject to the two-up one-down local constraint, in a similar sense to the two-in two-out state in spin ice~\cite{Harris1997,Ramirez1999}.
The spin state is composed of the emergent degrees of freedom, self-avoiding up-spin loops and isolated down-spins, as schematically shown in Fig.~\ref{fig:diagram}(b). 
Hence, we call this Coulombic state the loop liquid (LL). 

An interesting observation here is that the change between the FM, PFM, and LL states is smooth and there is no sign of phase transition. 
Both $m$ and $P$ changes continuously without showing any singularity, and the magnetic susceptibility $\chi_m$ shows only a broad hump, as shown in Fig.~\ref{fig:ndata}(b).  
This indicates that the change from FM to LL is a crossover and not a phase transition. 
Such behavior is understood from the symmetry point of view.
In the LL state, though $m$ is nonzero, the system still remains disordered and preserves all the symmetries of the lattice; the situation is unchanged from the FM and PFM states.
As a consequence, these phases are smoothly connected by the crossover.

On the other hand, as decreasing $T$ or as further increasing $n$, the LL state exhibits phase transitions with showing a magnetic long-range order (LRO). 
In our MC simulation, we identify two different transitions; one is the transition to the state with $q=0$ LRO of the two-up one-down spin configurations [Fig.~\ref{fig:diagram}(c)], and the other to the state with $\sqrt3\times\sqrt3$ LRO [Fig.~\ref{fig:diagram}(d)]. 
The former is observed while decreasing $T$ at $n\sim 0.83$, and the latter is found by increasing $n$ to a commensurate filling $n=8/9$.
$S(\bf{k})$ for the latter state is shown in Fig.~\ref{fig:ndata}(e). 
In the corresponding density regions, the two phases are obtained in the variational calculation for the ground state, as shown in Fig.~\ref{fig:diagram}(a). 
These two LRO states are viewed as crystal phases of the emergent loops in the two extreme cases; the former is a periodic array of one-dimensional chains, while the latter the shortest six-site hexagons.
Interestingly, the peculiar LL state extends in the density region between these two crystal phases.

\begin{figure}
   \includegraphics[width=\linewidth]{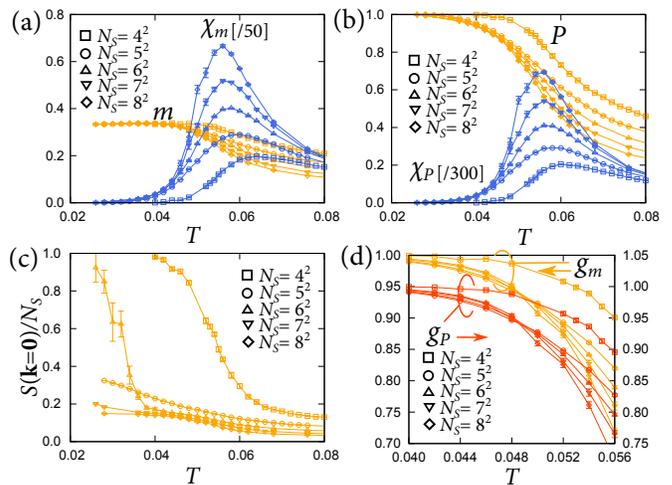}
   \caption{(color online).
   MC results for (a) $m$ and $\chi_m$, (b) $P$ and $\chi_P$, (c) $S({\bf k}={\bf 0})/N_{\rm s}$, and (d) $g_m$ and $g_P$ for $N_{\rm s}=4^2$, $5^2$, $6^2$, $7^2$, and $8^2$ and at $n=0.83$.
  }
   \label{fig:tdata}
\end{figure}

Let us closely look at the formation of LL and the crystallization of loops.
Figure~\ref{fig:tdata} shows the MC results of $T$ dependences of magnetic properties at $n = 0.83$. 
The result in Fig.~\ref{fig:tdata}(a) shows the increase of $m$ with saturation to $1/3$ and a divergent peak of $\chi_m$ at $T\sim 0.05$. 
At the same time, as shown in Fig.~\ref{fig:tdata}(b), $P$ shows saturation to 1 and its susceptibility $\chi_P$ shows a peak, indicating that most of the triangles become two-up one-down below $T\sim0.05$.
The Binder parameters~\cite{Binder1981} for $m$ and $P$, $g_m$ and $g_P$, respectively, are shown in Fig.~\ref{fig:tdata}(d).
Both show a crossing of the results for different sizes, indicating the transition is of second order.
The critical temperatures determined from the two independent Binder analyses show good accordance; $T_c=0.051(4)$.
On the other hand, a rapid increase of $S({\bf k}={\bf 0})/N_{\rm s}$ to 1 is observed in $N_{\rm s}=4^2$ and $6^2$, as shown in Fig.~\ref{fig:tdata}(c); 
the onset $T$ decreases for larger $N_{\rm s}$ although the results show strong finite size effects with different behavior for even and odd $N_{\rm s}$.
This suggests a phase transition to the $q=0$ ordered state at a lower $T$ than $0.028$; this is consistent with the ground state obtained by the variational calculation, as shown in Fig.~\ref{fig:diagram}(a).
Although the precise estimate of the critical temperature is difficult within the present calculation, these results indicate that successive phase transitions from PM to LL and LL to $q=0$ FR state take place at $n=0.83$.
The former corresponds to the formation of loops, and the latter their crystallization.

\begin{figure}
   \includegraphics[width=\linewidth]{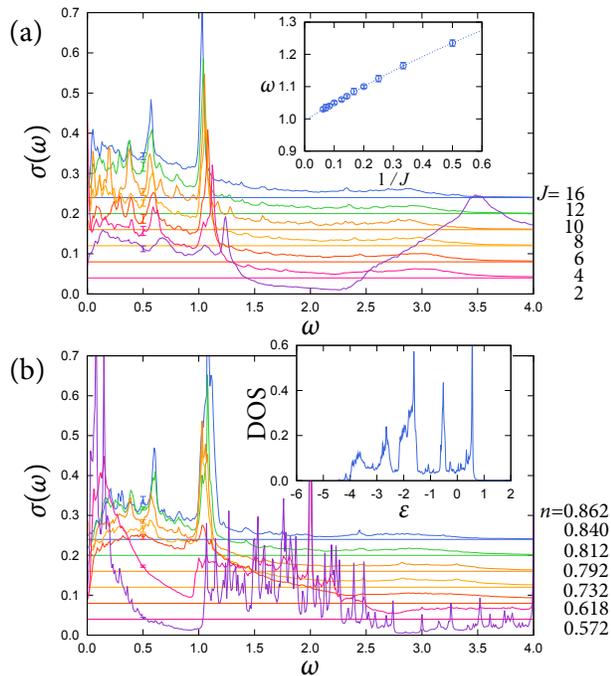}
   \caption{(color online).
   Optical conductivity $\sigma(\omega)$ calculated (a) by simple average over LL configurations while varying $J$ at $n=0.843$ for a $2^2$ supercell of $N=3\times 12^2$ sites, and (b) by MC simulation while varying $n$ at $J=6$ for a $4^2$ supercell of $N=3\times 6^2$ sites at $T=0.04$.
   The scattering rate in the Kubo formula is taken as $\tau^{-1}=0.01$.
   The typical error bars are shown at $\omega=0.5$.
   The inset in (a) shows the peak position of $\sigma(\omega)$ at $\omega \sim 1$. 
   The dotted line shows the fitting by $\omega=0.995+0.558/J-0.155/J^2$.
   The inset in (b) is DOS at $n=0.862$.
   The Fermi level is set at $\varepsilon=0$.
   }
   \label{fig:sigma}
\end{figure}

Now we discuss the electronic and transport properties of the itinerant electrons.
Figure~\ref{fig:sigma} shows the result of optical conductivity $\sigma(\omega)$.
First, to extract the effect of characteristic spin correlations in the LL state, we calculate $\sigma(\omega)$ by taking simple average over different spin patterns in the ideal LL manifold (all the triangles satisfy the two-up one-down local constraint).
The calculations were done by using the Kubo formula for 24 different spin patterns.
Figure~\ref{fig:sigma}(a) is the result of $\sigma(\omega)$ calculated at $n = 0.843$ for various $J$.
All the results show a sharp peak at $\omega=\omega_p\sim 1.0$-$1.2$, which shifts to lower $\omega$ for larger $J$.

The characteristic peak comes from the transition process between two localized states in the six-site loops.
In the limit of $J\to\infty$, electrons are confined in the loops or at isolated sites~\cite{Jaubert2012}; the contribution to $\sigma(\omega)$ comes only from the transition process between the electronic states in the same loop.
Hence, sharp peaks appear in $\sigma(\omega)$ corresponding to the discrete energy levels in the finite length loops.
In the current kagome case, the most dominant loops are the shortest ones with the length of six sites.
In the six-site loops, the energy difference between the unoccupied and occupied levels at this filling (the highest and second highest levels) is $1$.
Hence, we expect a sharp peak at $\omega_p=1$ in the limit of $J \to \infty$. 
For large but finite $J$, the second order perturbation in terms of the hopping between up and down spin sites shifts the second highest eigenenergy to a lower energy.
On the other hand, this perturbation process does not affect the highest eigenenergy. 
Hence, it is expected that the peak shifts to a higher $\omega$ as decreasing $J$; the asymptotic behavior at $J\to \infty$ is expected to be $\omega_p = 1+{\cal O}(1/J)$.
This is confirmed by the fitting shown in the inset of Fig.~\ref{fig:sigma}(a).

Interestingly, the peak persists in the weak $J$ region where the exchange splitting $2J$ is comparable or smaller than the bare bandwidth $6t$ and the above perturbative argument appears to be no longer valid.
In a recent study on a metal-insulator transition caused by correlated potentials, a LL-type local correlation induces a metal-insulator transition at a considerably smaller potential than the bandwidth by confining the electrons in the loops~\cite{Ishizuka2011}. 
The persisting resonant peak in $\sigma(\omega)$ is likely to be the consequence of this confinement.

Emergence of the characteristic peak is also observed in the thermodynamic average obtained by the MC simulation.
Figure~\ref{fig:sigma}(b) shows the MC result of $\sigma(\omega)$ while varying $n$ at $T=0.04$ and $J=6$.
With increasing $n$ from the FM region, the peak at $\omega\sim 1$ develops in the LL state for $n \gtrsim 0.8$.
The inset in Fig.~\ref{fig:sigma}(b) shows the density of states (DOS) for itinerant electrons (lower half of two split bands) at $n=0.862$.
The result clearly shows the presence of two sharp peaks below and above the Fermi level set at $\varepsilon=0$; the energy difference is about $1.1$, which well corresponds to the peak in $\sigma(\omega)$ in the main panel of Fig.~\ref{fig:sigma}(b).

To summarize, we studied an Ising-spin Kondo lattice model on a kagome lattice with focusing on the emergent magnetic states and their electronic properties.
By using an unbiased Monte Carlo simulation, we presented that the loop-liquid state emerges in the finite temperature region, in addition to ferromagnetic, $q=0$ ferrimagnetic, and $\sqrt3\times\sqrt3$ ferrimagnetic states.
The loop liquid is a Coulombic ferrimagnetic state, characterized by the emergent up-spin loops originating from the two-up one-down local spin configurations.
The phase diagram is understood in terms of the emergent loops as crystallization and cohesion of the dense liquid of the loops.
We also showed that the loop-liquid formation is observed in characteristic peaks in the optical conductivity.
Recently, the spin-charge coupling in frustrated magnets has been revealed to exhibit rich physics, both in magnetic and transport properties. 
We hope that our finding of yet another emergent state would further stimulate the study of these systems.

The authors thank N. Furukawa, L. D. C. Jaubert, and K. Penc for fruitful discussions.
H.I. is supported by Grant-in-Aid for JSPS Fellows.
This research was supported by KAKENHI (No.~22540372 and 24340076), the Strategic Programs for Innovative Research (SPIRE), MEXT, and the Computational Materials Science Initiative (CMSI), Japan.


\begin{thebibliography}{99}
\bibitem{note_pfm}          At the lowest $T$, the PFM phase is presumably taken over by a phase separation between FM and $q=0$ FR states [see Fig.~\ref{fig:diagram}(a)].
\end{thebibliography}
\end{document}